# Synthetic vs human emotional faces: what changes in humans' decoding accuracy.

T. Amorese, M. Cuciniello, A. Vinciarelli, G. Cordasco, A. Esposito

*Abstract*— **Considered the increasing use of assistive technologies in the shape of virtual agents, it is necessary to investigate those factors which characterize and affect the interaction between the user and the agent, among these emerges the way in which people interpret and decode synthetic emotions, i. e. emotional expressions conveyed by virtual agents. For these reasons, a study is proposed, in which were involved 278 participants split in differently aged groups (young, middle-aged and elders). Within each age group some participants were administered a "naturalistic decoding task", a recognition task of human emotional faces while other were administered a "synthetic decoding task" namely emotional expressions conveyed by virtual agents. Participants were required to label pictures of female and male humans or virtual agents of different ages (young, middle-aged and old) displaying static expressions of disgust, anger, sadness, fear, happiness, surprise, and neutrality. Results showed that young participants showed better recognition performances (compared to older groups) of anger, sadness, and neutrality while female participants showed better recognition performances (compared to males) of sadness, fear, and neutrality; sadness and fear were better recognized when conveyed by real human faces, while happiness, surprise and neutrality were better recognized when represented by virtual agents. Young faces were better decoded when expressing anger and surprise, middle-aged faces were better decoded when expressing sadness, fear, and happiness while old faces were better decoded in the case of disgust; on average female faces where better decoded compared to male ones.**

## I. Introduction

Emotions play a central role, along with human life, also in human-machine interaction processes [1]. To make the interaction between humans and machines, such as virtual agents, social robots and chatbots, more effective, it is necessary not only to equip them with the ability to appropriately respond to human emotions, but also with the ability to manifest emotional expressions [2], since this skill seems to have a strong impact on users, influencing their behavior and attitudes [3]. A fundamental aspect is certainly the ability to interpret and decode emotions, especially when a person is interacting with assistive technologies such as conversational agents, capable of manifesting emotional expressions through the face and / or the voice. Nevertheless, emotion recognition is a complex process which could be affected by several variables, both inherent to the person decoding emotional expressions (such as age and gender) and the features of the face or the voice to be decoded (age, gender, typology, etc.).

The possibility to exploit virtual agents able to manifest emotional expressions is fundamental since, as already shown [4], people prefer to communicate with virtual agents able to convey emotional facial expressions rather than with virtual agents "emotion-less", although the expressions are subtle and not particularly marked. Assumed that people seem to be capable to correctly interpret affective meanings conveyed by emotional virtual characters [5], studies investigating and comparing peoples' ability to decode emotional expressions conveyed by human beings and virtual agents led to different and conflicting results, some outlining that people do not show differences while recognizing emotions expressed by real human and virtual agents' faces [6][7][8], while other researches highlighted that people tend to better decode emotions conveyed by humans rather than virtual agents [9][10][11][12][13]. It also emerged that results could depend on the emotional category considered, as showed by studies highlighting that disgust, is better decoded when conveyed by human beings while sadness and fear are better decoded when conveyed by virtual agents rather than human faces [14][15]. Another study [16] showed people higher percentage accuracy in the identification of emotions displayed by dynamic virtual agents rather than static human faces.

In addition to the typology of face showed, for instance human or non-human, the age of a face is an additional variable which could affect people's ability to recognize emotions. According to a theoretical model known as the "Own Age Bias Theory", people better recognize facial expressions and emotions displayed by peers than those expressed by people belonging to different age groups. While some studies have shown that this

T. Amorese, M. Cuciniello, G. Cordasco and A. Esposito are with department of psychology at Università degli Studi della Campania "L. Vanvitelli", Caserta, Italy, and The International Institute for Advanced Scientific Studies (IIASS), Vietri sul Mare, Italy. A. Vinciarelli is with School of Computing Science and the Institute of Neuroscience and Psychology, University of Glasgow, Glasgow, UK. (email: terry.amorese@unicampania.it; marialucia.cuciniello@unicampania.it; gennaro.cordasco@unicampania.it; anna.esposito@unicampania.it; Alessandro.Vinciarelli@glasgow.ac.uk ).

The research leading to these results has received funding from the European Union Horizon 2020 research and innovation program under grant agreement N. 769872 (EMPATHIC) and N. 823907 (MENHIR), from the project SIROBOTICS that received funding from Ministero dell'Istruzione, dell'Università, e della Ricerca (MIUR), PNR 2015-2020, Decreto Direttoriale 1735 July 13 2017, and from the project ANDROIDS that received funding from Università della Campania "Luigi Vanvitelli" inside the program V:ALERE 2019, funded with D.R. 906 del 4/10/2019, prot. n. 157264, October 17, 2019.

is true for adults [17][18], who tend to recognize facial expressions of peers more quickly and accurately than those expressed by younger or older people, results for children and adolescents are conflicting. In fact, while some studies have shown the existence of the Own Age Bias in children, who tend to better recognize facial expressions of their peers [19], others showed that children can recognize emotional expressions equally well, regardless of whether these are expressed by their peers or by adults [20]. Even the gender of the face which expresses an emotion could have an impact on facial expressions' accuracy recognition. A theory known as the "Own Gender Bias" states that people could recognize more quickly and easier faces belonging to their own gender compared to faces and emotional expressions of the other gender, in other words men are better at decoding male facial expressions and women are better at decoding female facial expressions [21][22].

Other than the features of the face to be decoded, some characteristics of the person who is decoding an emotional expression could also have an effect on this process, among these, age emerges. For instance, it seems that aging is associated with a decreased ability to recognize specific emotional expressions. Studies involving people belonging to different age groups showed less accuracy in the recognition of emotions such as fear, anger, and sadness by middle-aged and elderly people [23][24][25][26] compared to younger participants. Several studies compared young and older people's ability to recognize emotional expressions, showing better performances of younger participants compared to elderly, especially concerning recognition of specific emotions such as anger and sadness [27][28]; moreover, similar difficulties are also evident as regards the vocal expression of these emotions [29]. Even the gender of the person who is interpreting an emotion is a factor that could influence this process, in fact differences between men and women concerning the ability to decode emotions is a widely investigated topic; some studies highlighted women's greater ability to accurately recognize emotional expressions compared to men [30]31][32]. However, other studies have shown that the effect of the gender could be mediated by other factors such as the intensity of the stimulus, highlighting that, women would be more accurate than men in recognizing only subtle emotional expressions [33]. Other research has shown that the modality of presentation of the stimulus and the sensory channel involved could also influence the ability to accurately decode emotions, more specifically women seem to be more accurate than man in decoding emotional prosody [34].

Considered all the above-mentioned factors, the impact that these could have on emotion recognition abilities, and the relevance of emotions within human-machine interaction, a study is proposed. Static pictures depicting humans and virtual agents were exploited within this research in which differently aged group of participants were required to decode expressions of disgust, anger, sadness, fear, happiness, surprise, and neutrality.

## II. Materials and Methods

The present investigation aims at assessing:

- The effect of participants' age (young, middle-aged, and old) and gender on the ability to decode static facial emotional expressions.
- The effect of stimuli's age (young, middle-aged, and old) and gender on participants' recognition scores of facial emotional expressions.
- Differences in the recognition of static facial expressions conveyed by human beings (naturalistic) and by virtual agents (synthetic).

### A. *Participants*

The experiments involved a total of 278 differently aged healthy subjects split into six groups: 45 young participants aged between 22 and 35 years (mean age=27.33; SD= ±4.10, 23 females); 45 middle-aged participants between 40 and 55 years (mean age=48.07; SD= ± 5.76, 24 females); 45 elders aged 65+ (mean age=73.60; SD=±7.48, 22 females), which were administered a "naturalistic decoding task" namely a recognition task of human emotional faces taken from popular movies (SFEW database). 50 young participants aged between 22 and 35 years (mean age=27.86; SD= ±2.75, 30 females); 48 middle-aged participants between 40 and 55 years (mean age=50; SD= ± 4.14, 26 females); 45 elders aged 65+ (mean age=66; SD=±1.41, 25 females), which were administered a "synthetic decoding task" namely emotional expressions conveyed by virtual agents. Participants were all Italians living in the Campania region. Participants which were administered the naturalistic task were recruited with a face-to-face modality, while, considered the rise of the Covid-19 pandemic disease and the need to respect social distancing, participants which were administered the synthetic task were recruited exploiting telematic tools as social networks and e-mails; they voluntarily joined the study and read and agreed to an informed consent formulated according to the Italian and European laws about privacy and data protection. The research was authorized by the ethical committee of the Department of Psychology at the Università degli Studi della Campania "Luigi Vanvitelli" with the protocol number 25/2017.

### B. *Stimuli*

The naturalistic task required participants to decode pictures depicting differently aged (young, middle- aged and old) female and male actors expressing disgust, anger, sadness, fear, happiness, surprise, and neutrality. Pictures were taken from the Static Facial Expressions in the Wild (SFEW) database, developed, and validated by Dhall and colleagues [35] by selecting frames from the Acted Facial Expressions in the Wild (AFEW), which consists of dynamic emotional faces extracted from movies. The database contains faces of actors aged from 1 to 70 years and consists of unconstrained emotional faces varied head poses. SFEW contains 700 emotional faces expressing the six basic emotions of anger, disgust, fear, happiness, sadness, surprise, and neutrality. The emotional labels were assigned to the faces by two independent labelers. The current experiment exploits 42 faces of young, middle-aged, and old women and men displaying the seven facial expressions of disgust, anger, fear, sadness, happiness, surprise, and neutrality. In order to equally balance male and female faces, the SFEW stimuli were supplemented with seven more of: a) one old female and one young male faces of disgust, b) one old female face of sadness, c) one old female face of neutrality, d) one old female and a middle-aged female faces of fear, c) one old female face of surprise. These stimuli were obtained from pictures available on internet. Fig.1 illustrates some examples of stimuli used within the naturalistic experiment.

The synthetic experiment consisted in an emotion decoding task in which participants were showed pictures depicting differently aged (young, middle-aged, and old) female and male virtual agents expressing disgust, anger, sadness, fear, happiness, surprise, and neutrality. Virtual agents were developed in collaboration with Paphus Solutions Inc., a Canadian corporation specialized in the development of bots, artificial intelligence, and deep learning products and services. The proposed virtual agents were developed using Daz3D software which provides emotion pose presets for most emotions. For each emotion, the expression morph from Daz3D has been used at the 100% magnitude (intensity of the emotional expression) setting. For the current experiment, 42 pictures depicting virtual agents were exploited. Stimuli of both the naturalistic and the synthetic tasks were validated in a qualitative manner by three experts in the field of emotional interactional exchanges and human-machine interaction; raters always agreed on stimuli's ability to convey a specific emotion, highlighting an inter-rater confidence rate of 100%. Fig.2 shows some examples of pictures depicting virtual agents exploited within the synthetic experiment. The faces depicted were all Caucasian, although this was a coincidence, as faces' ethnical features were not a variable of interest in our study.

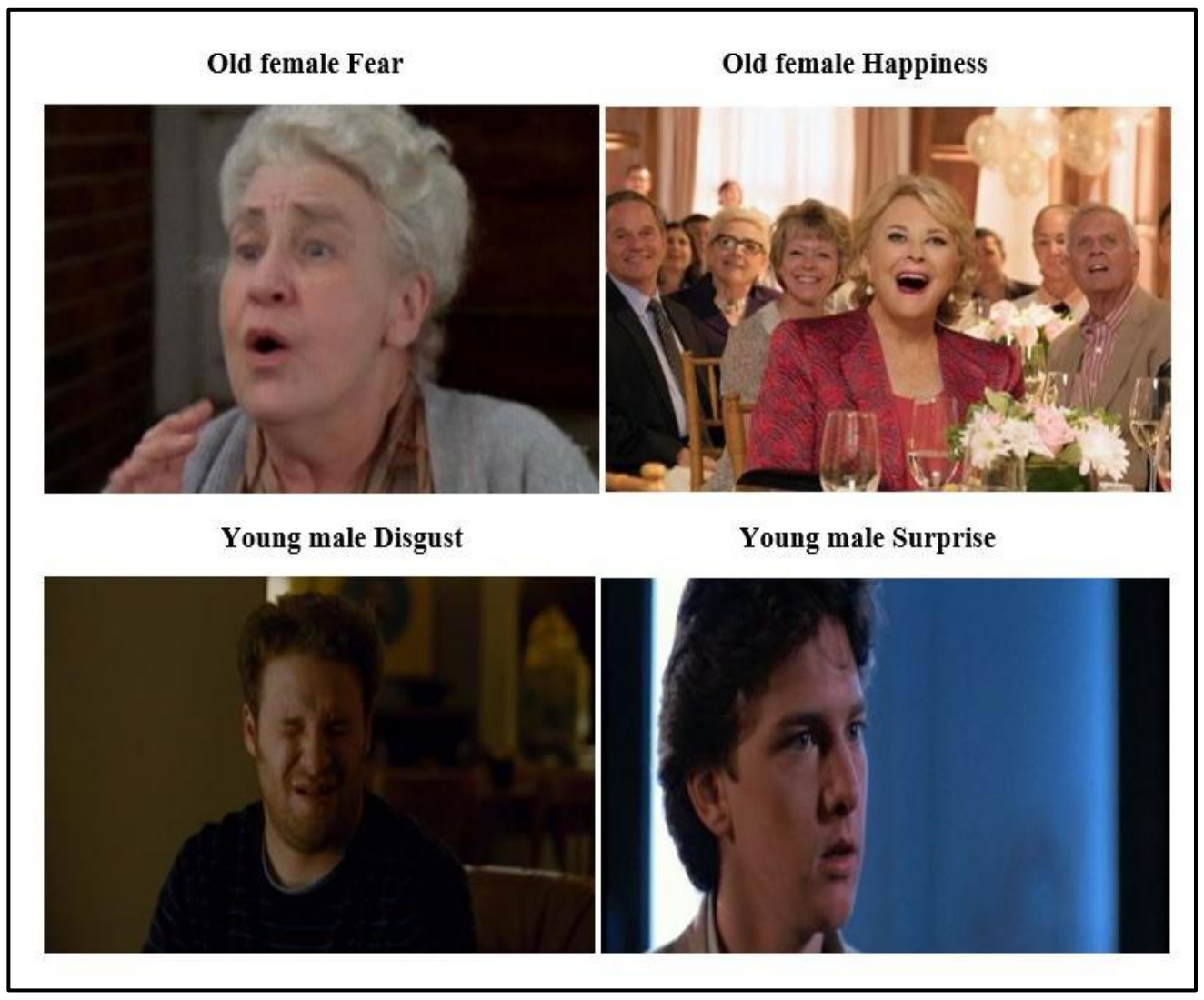

**Fig. 1.** A sample of stimuli exploited for the naturalistic decoding task.

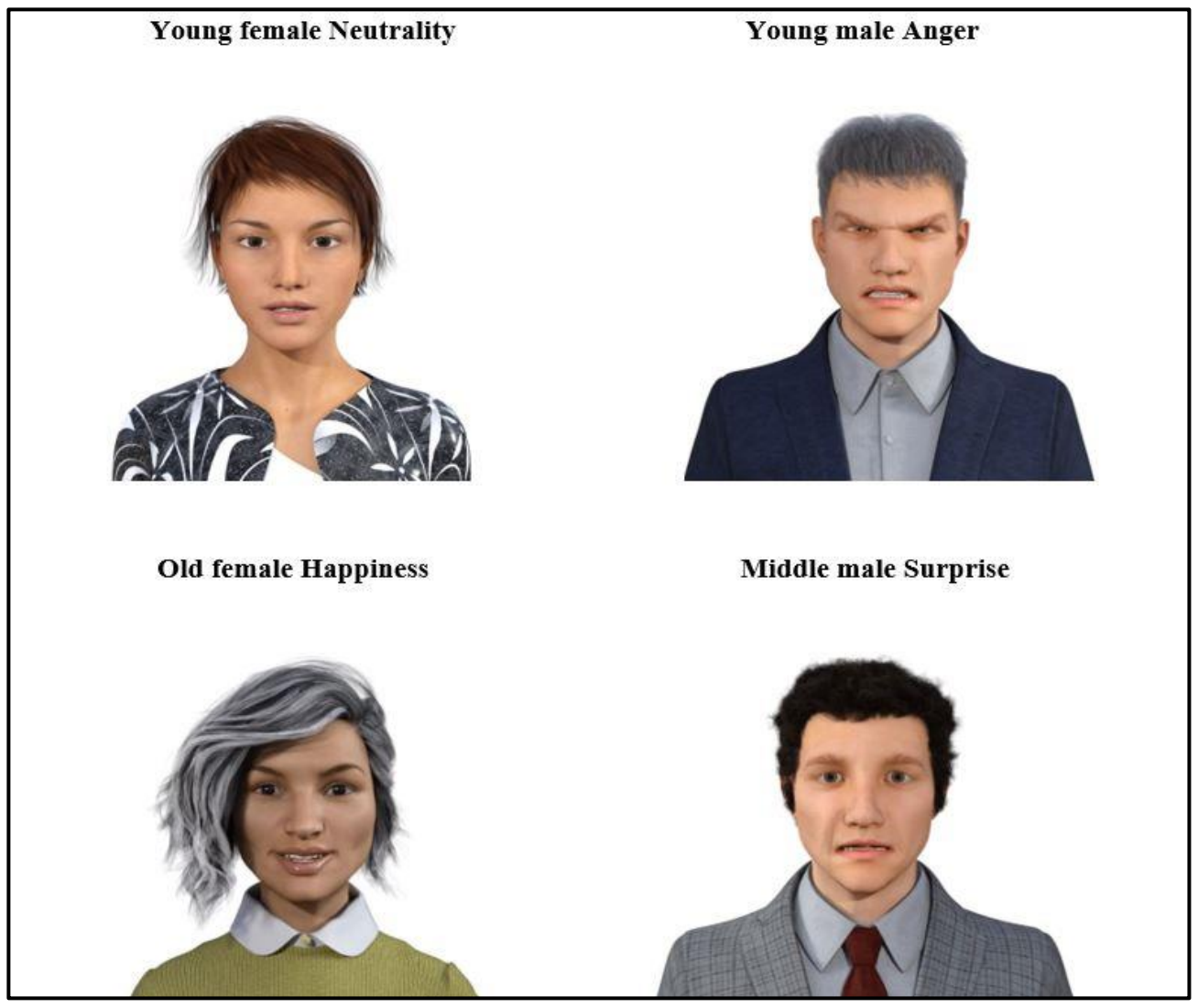

**Fig. 2**. A sample of stimuli exploited for the synthetic decoding task.

### *C. Tools and Procedures*

The naturalistic experiment consisted in an emotion recognition task developed using the SuperLab 4.0 software. Participants read and signed an informed consent, then they were asked to sit in front of a laptop, on which the experiment was running in order to accomplish the facial expression decoding task. The task consisted of a trial session where 3 randomized static pictures were presented, allowing participants to familiarize with the decoding procedure. Then the experimental session consisting in the presentation of 42 randomized static pictures was proposed. After viewing each stimulus, participants had to attach an emotional label to it, pressing a laptop key corresponding to the envisaged emotion: d= disgust, a=anger, f=fear, s=sadness, h=happiness, s=surprise and n=neutrality. Since the instructions were presented only once before running the experiment, a small legend describing the laptop keys corresponding to the emotional label was placed next to the laptop monitor to act as reminder. The synthetic experiment was developed using the online study builder Lab.js and then exported on JATOS (Just Another Tool for Online Studies), which allows to generate links which can be sent to participants. Volunteers were recruited exploiting social networks and e-mail addresses.

Each participant who agreed to join the study was provided with a link and asked to open the link from a laptop. Once opened the link, participants were asked to give their consent to a personal data processing form, subsequently participants' demographic data and information about their degree of experience with technology were collected. After this, on the screen appeared the instructions required to carry out the experiment, followed by a trial session (composed by 3 static pictures) and by an experimental session (composed by 42 static pictures). Both the trial and the experimental session consisted in the presentation of randomized stimuli. For each stimulus, participants had to attach an emotional label choosing among these options: disgust, anger, sadness, fear, happiness, surprise, neutrality, or other emotion. For both the naturalistic and the synthetic experiments when, according to the participant, none of the categories described the emotion seen, they could choice the "other emotion" option of response. We included the response option "other emotion" to give participants greater freedom of response, but since the pictures ware necessarily associated to an emotional category, when the participant chose this option, the result is equivalent to not having correctly decoded the emotion, so it is actually contained in the results as an incorrect answer. Moreover, considering that each emotional category was represented by one female and one male young face, one female and one male middle-aged face and one female and one male old face, in both experiments each emotion was administered six times.

## III. Data analysis

Repeated measures ANOVA were performed on the collected data. The significance level was set at $\alpha < .05$ and differences among means were assessed through Bonferroni's post hoc tests. These analyses are discussed in detail in the Appendix A. To test if the preconditions to run ANOVA were met, normality tests were carried out, highlighting that some emotional categories are normally distributed, and others are not; however, the reported significances are the same for both parametric and non-parametric tests. So, we preferred to use ANOVA for consistency and also because, participant sample was pretty large (278) and according to the "Central Limit Theorem" when independent random variables are summed up, their sum tends toward a normal distribution, in particular when the sample is large and because ANOVA is pretty robust to non-normality.

## IV. Results

### A. *Effects of participants gender, age, and type of stimulus on decoding accuracy of emotional expressions.*

Table 1 and 2 displays the decoding accuracy (in %) over all the proposed emotional categories and Figure 3 show mean decoding differences among the differently aged participants and for each emotion.

**Table 1**: All participants' decoding accuracy (in %) of naturalistic and synthetic emotional faces

| ACCURACY % | *Disgust* | *Anger* | *Sadness* | *Fear* | *Happiness* | *Surprise* | *Neutral* |
|---|---|---|---|---|---|---|---|
| **NATURALISTIC FACES** | 25,80 | 71,11 | 50,74 | 23,83 | 60,74 | 36,30 | 39,63 |
| **SYNTHETIC FACES** | 28,32 | 73,08 | 41,03 | 15,27 | 70,28 | 46,27 | 61,19 |

**Table 2**: Young, Middle- aged and Elders decoding accuracy (in %) of naturalistic and synthetic emotional faces.

| YOUNG ACCURACY % | *Disgust* | *Anger* | *Sadness* | *Fear* | *Happiness* | *Surprise* | *Neutral* |
|---|---|---|---|---|---|---|---|
| **NATURALISTIC FACES** | 34,81 | 75,93 | 57,78 | 27,04 | 61,48 | 37,41 | 45,93 |
| **SYNTHETIC FACES** | 28,67 | 80,00 | 44,67 | 16,33 | 74,33 | 55,33 | 73,00 |
| MIDDLE-AGED ACCURACY % | *Disgust* | *Anger* | *Sadness* | *Fear* | *Happiness* | *Surprise* | *Neutral* |
| **NATURALISTIC FACES** | 18,52 | 66,30 | 49,63 | 19,63 | 56,67 | 41,11 | 31,11 |
| **SYNTHETIC FACES** | 31,60 | 72,92 | 40,63 | 15,28 | 65,63 | 48,61 | 58,68 |
| ELDERS ACCURACY % | *Disgust* | *Anger* | *Sadness* | *Fear* | *Happiness* | *Surprise* | *Neutral* |
| **NATURALISTIC FACES** | 24,07 | 71,11 | 44,81 | 24,81 | 64,07 | 30,37 | 41,85 |
| **SYNTHETIC FACES** | 24,44 | 65,56 | 37,41 | 14,07 | 70,74 | 33,70 | 50,74 |

Table 1 highlights that Sadness and Fear were more accurately decoded on naturalistic faces, while Disgust, Anger, Happiness, Surprise and Neutrality were decoded more accurately on synthetic faces. As showed by table 2, young participants who were administered naturalistic faces decoded Disgust, Sadness and Fear more accurately compared to young administered with synthetic faces, while Anger, Happiness, Surprise and Neutrality were better decoded by young administered synthetic facial expressions. Middle-aged participants who were administered naturalistic faces decoded Sadness and Fear more accurately compared to middle-aged administered with synthetic faces, while Disgust, Anger, Happiness, Surprise and Neutrality were better decoded by the group of middle- aged administered synthetic facial expressions. Elders administered naturalistic faces decoded Anger, Sadness and Fear more accurately compared to elders administered with synthetic faces, while Disgust, Happiness, Surprise and Neutrality were more accurately decoded by elders administered synthetic facial expressions. The graph below shows that young participants displayed better decoding performances compared to middle-aged and old participants, synthetic facial expressions were better decoded compared to naturalistic ones; anger and happiness were the emotional categories better decoded among those proposed.

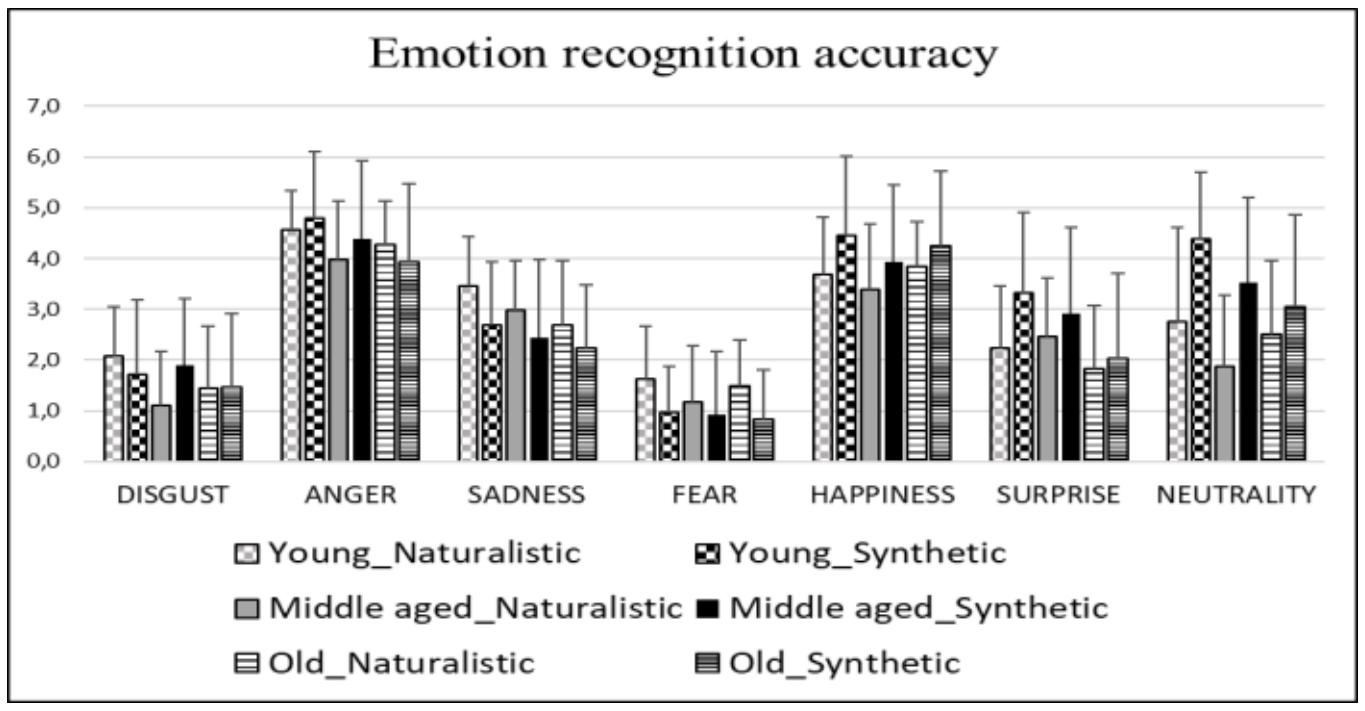

**Fig. 3**. Recognition scores of each age group administered naturalistic and synthetic faces for each emotional category. Y- axis represent mean score which can vary between 0 and 6 since it represents total decoding scores for each emotional category (the total score is the sum of the scores obtained by each stimulus, there are 6 stimuli for each emotion, so means vary from 0 to 6).

### B. *Effects of participants' and stimuli's age and gender on the decoding accuracy of naturalistic and synthetic emotional faces.*

Table 3 summarize factors affecting participants' decoding accuracy of naturalistic and synthetic faces, these analyses are discussed in detail in Appendix A, section A.2. As showed in table 3 and figure 4, according to the different emotional category considered, each investigated variable exerts different effects on the obtained results. **Disgust** was better decoded by young participants administered naturalistic faces compared to middle-aged participants; middle-aged participants administered synthetic facial expressions showed better recognition performances compared to the group of middle - aged required to decode naturalistic faces. Old faces of disgust were better recognized compared to young facial faces and female facial expressions were better decoded compared to male facial expressions. When decoding **anger**, young participants showed better recognition performances compared to middle-aged and old participants; young faces were better recognized compared to middle-aged and old ones. **Sadness** was better decoded by female participants rather than male participants, moreover young participants showed better decoding performances compared to old participants. Results also showed that sad faces were better decoded when conveyed by human faces compared to virtual agents' faces and that middle-aged facial expressions of sadness were better recognized compared to young and old facial expressions; even in this case female facial expressions were better recognized compared to male facial expression.

**Fear** was better decoded by female participants compared to males and when conveyed by real human faces rather than virtual agents' ones. Middle-aged facial expressions of fear were better recognized compared to young and old facial expressions; female facial expressions were better recognized compared to male facial expressions.

**Table 3**: Summary of factors affecting participants' decoding accuracy of emotional faces.

| | **Effects of participants' gender** | **Effects of participants' age** | **Effects of type of stimuli: Naturalistic vs Synthetic** | **Gender of stimuli effect** | **Age of stimuli effect** | **Significant interactions** |
|---|---|---|---|---|---|---|
| **DISGUST** | No significant participants' gender effects. | Bonferroni's post hoc multiple adjustments do not allow to identify which means produce differences. | No significant type of stimuli effects. | Female significantly more accurately decoded than male faces. | Old more accurately decoded than young faces. | 1) Participants' age and the type of stimulus<br>2)Age and gender of stimuli<br>3) Age and type of stimuli<br>4) Participants' age and gender of stimuli |
| **ANGER** | No significant participants' gender effects. | Young participants showed better decoding performances compared to middle-aged and old participants. | No significant type of stimuli effects. | Female better recognized compared to male faces. | Young better recognized compared to middle-aged and old faces. | 1) Age and gender of stimuli<br>2) Age and type of stimuli<br>3) Type and gender of stimuli |
| **SADNESS** | Female participants better performances than males' ones. | Young better decoding performances compared to old participants. | Naturalistic faces better decoded compared to synthetic ones. | Female better recognized compared to male faces. | Middle-aged better recognized compared to young and old faces. | 1) Age and gender of stimuli<br>2) Age and type of stimuli<br>3) Type and gender of stimuli |
| **FEAR** | Female participants better performances than males' ones. | No significant participants' age effects. | Naturalistic faces better decoded compared to synthetic ones. | Female better recognized compared to male faces. | Middle-aged better recognized compared to young and old faces. | 1) Age and gender of stimuli<br>2) Age and type of stimuli<br>3) Participants' and stimuli's gender<br>4) Participants' age and gender of stimuli |
| **HAPPINESS** | No significant participants' gender effects. | No significant participants' age effects. | Synthetic faces better decoded compared to naturalistic ones. | Female better recognized compared to male faces. | Middle-aged better recognized compared to young and old faces. | 1) Age and gender of stimuli<br>2) Participants'age and age of stimuli<br>3) Age and type of stimuli<br>4) Type and gender of stimuli |
| **SURPRISE** | No significant participants' gender effects. | Elders showed worse decoding performances compared to middle-aged and young participants. | Synthetic faces better decoded compared to naturalistic ones. | No significant gender of stimuli effects. | Young better recognized compared to middle-aged and old faces. | 1) Participants'age and gender<br>2) Participants'age and age of stimuli<br>3) Type and age of stimuli<br>4) Type and gender of stimuli<br>5) Age and gender of stimuli |
| **NEUTRALITY** | Female participants better performances than males' ones. | Young participants showed better decoding performances compared to middle-aged and old participants. | Synthetic faces better decoded compared to naturalistic ones. | Male significantly more accurately decoded than female faces. | No significant age of stimuli effects. | 1) Participants' age and stimulus type<br>2) Type and age of stimuli<br>3) Age and gender of stimuli |

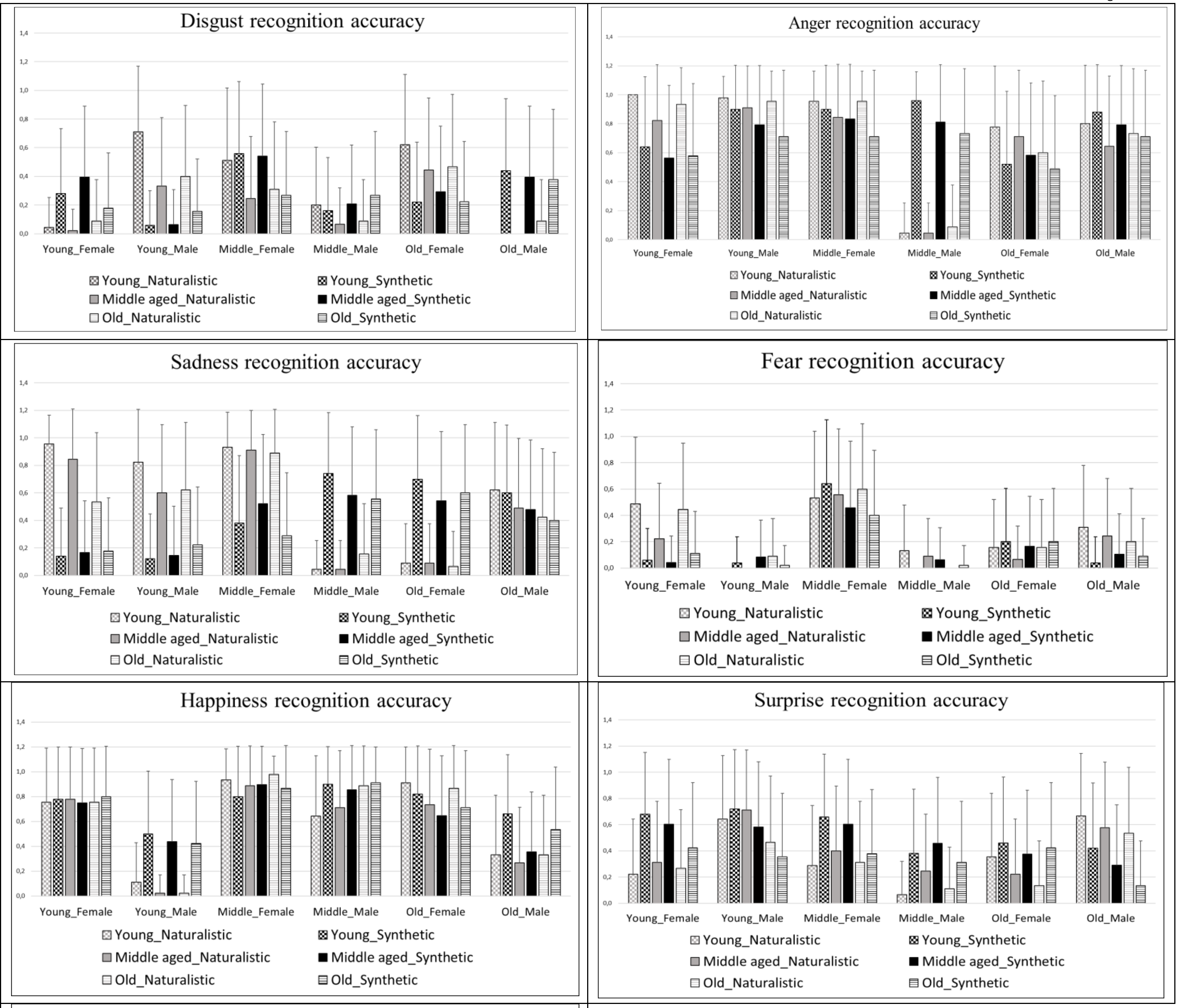

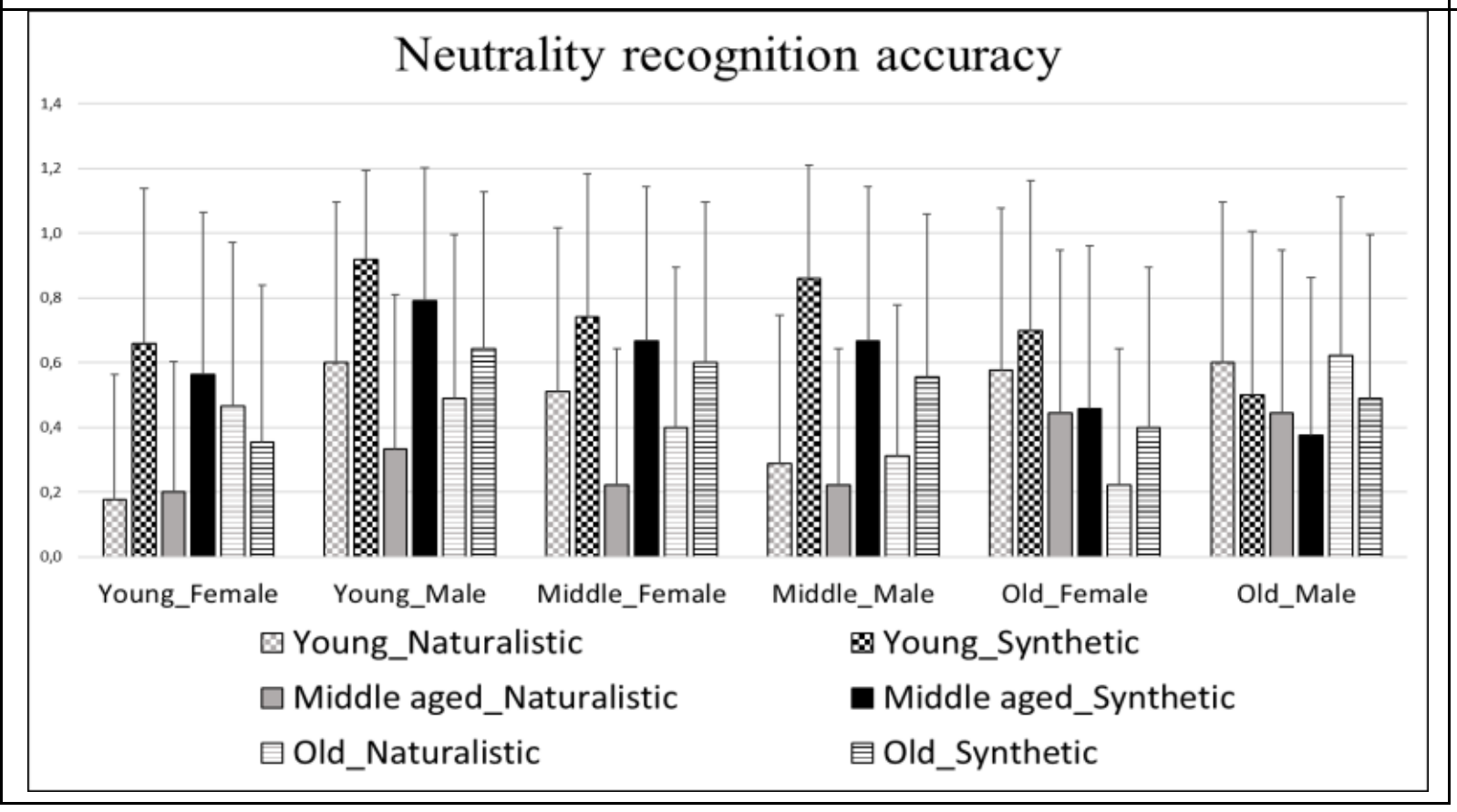

**Fig. 4.** Recognition scores for each emotional categories conveyed by humans (naturalistic) and virtual agents (synthetic); means are showed for each stimulus. Y- axis represent mean score which can vary between 0 and 1 since when participants correctly decoded emotions, the attributed score was equal to 1, when the emotion was not correctly decoded the attributed score was equal to 0. Bars correspond to mean scores, which can vary between 0 and 1 (where the bar is missing it indicate that the average value is equal to 0).

**Happiness** was better recognized when conveyed by virtual agents rather than real human beings; middle-aged facial expressions of happiness were better recognized compared to young and old facial expressions. Female facial expressions of happiness were better decoded compared to male facial expressions. **Surprise** was better decoded by middle-aged and young participants compared to old ones and was better decoded when conveyed by virtual agents rather than by human beings. Young facial expressions of surprise were better recognized compared to middle-aged and old facial expressions. **Neutrality** was better decoded by female participants rather than males and by young participants compared to middle-aged and old ones. It was also observed that neutral expressions were better decoded when conveyed by virtual agents rather than human beings and that male facial expressions were better recognized compared to female facial expressions.

Table 4 show confusion matrices of participants' decoding accuracy (in %) for each emotional category for both the Naturalistic and Synthetic experiment. For disgust, naturalistic stimuli where mostly confused with anger, sadness, neutrality or identified as other emotion. Synthetic faces were mostly confused with anger. **Anger** was mostly confused with disgust, for both the naturalistic and the synthetic experiment. As regards as sadness, both naturalistic and synthetic faces of sadness were mostly confused with neutrality. Concerning **fear**, naturalistic and synthetic faces were mostly confused with surprise. For **happiness**, naturalistic facial expressions were mostly confused with surprise and neutrality, while synthetic faces were mostly confused with neutrality. For **surprise,** naturalistic stimuli were mostly confused with fear, happiness neutrality or identified as other emotion. Synthetic faces instead, were confused with fear. Concerning **neutrality,** naturalistic faces were confused with sadness, surprise, or identified as other emotion, synthetic ones were mostly confused with sadness.

**Tab.4**: Confusion matrices of participants decoding accuracy (in %) for each emotional category.

| **NATURALISTIC %** | **Disgust** | **Anger** | **Sadness** | **Fear** | **Happiness** | **Surprise** | **Neutral** |
|---|---|---|---|---|---|---|---|
| **Disgust** | **25,80** | 7,78 | 4,69 | 3,21 | 1,11 | 3,46 | 1,98 |
| **Anger** | 18,15 | **71,11** | 5,93 | 3,33 | 1,11 | 4,32 | 2,72 |
| **Sadness** | 11,73 | 3,09 | **50,74** | 12,10 | 3,33 | 9,14 | 16,91 |
| **Fear** | 7,28 | 4,57 | 3,70 | **23,83** | 2,96 | 11,85 | 3,33 |
| **Happiness** | 4,20 | 1,11 | 0,37 | 0,37 | **60,74** | 12,96 | 8,89 |
| **Surprise** | 9,14 | 5,31 | 10,37 | 29,38 | 10,49 | **36,30** | 10,00 |
| **Neutral** | 12,10 | 2,84 | 14,57 | 12,84 | 12,47 | 10,74 | **39,63** |
| **Other Emotion** | 11,60 | 4,20 | 9,63 | 14,94 | 7,78 | 11,23 | 16,54 |
| **SYNTHETIC%** | | | | | | | |
| **Disgust** | **28,32** | 13,87 | 12,59 | 3,38 | 0,93 | 4,55 | 2,68 |
| **Anger** | 59,21 | **73,08** | 5,94 | 3,50 | 0,58 | 3,26 | 3,15 |
| **Sadness** | 5,48 | 2,33 | **41,03** | 8,74 | 0,70 | 8,04 | 13,05 |
| **Fear** | 2,68 | 3,26 | 3,96 | **15,27** | 0,82 | 22,03 | 1,75 |
| **Happiness** | 0,00 | 0,35 | 0,58 | 10,37 | **70,28** | 2,33 | 3,15 |
| **Surprise** | 0,35 | 4,20 | 6,88 | 44,06 | 6,41 | **46,27** | 7,69 |
| **Neutral** | 1,98 | 2,10 | 20,75 | 9,21 | 14,92 | 9,09 | **61,19** |
| **Other Emotion** | 1,98 | 0,82 | 8,28 | 5,48 | 5,36 | 4,43 | 7,34 |

## V. DISCUSSION

All the investigated variables, more specifically participants' age and gender as well as the age, the gender and the type of stimulus administered seem to affect the way people decode emotional expressions, even though in a different way according to the emotional category considered. Participants' age had a significant impact on the study, as we found that young participants on average showed better recognition performances compared to participants belonging to older groups made up of middle-aged subjects between 40 and 55 years and seniors over 65, this is particularly true as regards as anger, sadness, surprise, and neutrality.

In literature aging effects on emotion decoding ability have been widely investigated, it seems that aging is associated with a decreased ability to recognize specific emotional expressions. As highlighted by the present research, studies involving people belonging to different age groups showed less accuracy in the recognition of emotions such as anger and sadness by middle-aged and elderly people [23][24][25][26][27][28] compared to younger participants. A possible explanation of difficulties showed by elderly in the recognition of specific emotional expressions regards the natural deterioration of those brain areas responsible of emotions' processing, as the frontal and medial temporal areas [36][37][38], this hypothesis is supported by lesions studies which demonstrated that patients with lesions in these areas show several difficulties in the recognition of facial expressions and emotions [39][40][41][42]. According to a different point of view elderly's difficulties in emotion recognition are mainly due to the type of task administered. In fact, differences between young and older participants seems to disappear when the emotion recognition task becomes "multimodal", namely when the emotion, is not expressed through a single modality, for example, only through facial expressions or only through the voice [43].

The present study also highlighted participants' gender effects on emotion recognition abilities, in particular as regards the recognition of sadness, fear and neutrality, with female participants showing greater decoding abilities compared to male ones.

Concerning the type of stimulus administered, discording results were observed depending on the emotional category considered, more in detail sadness and fear were better decoded by the group administered with real human faces, in contrast with results obtained in other studies in which instead sadness and fear seem to be better recognized when represented by synthetic faces [15,44,45]. Our study highlighted that happiness, surprise and neutrality were better recognized by the groups administered with emotional virtual agents. Moreover, while in the present study where not found differences between the group who saw synthetic and naturalistic angry faces, in literature positions concerning anger recognition are conflicting, since some studies showed that this emotion is better recognized when expressed by synthetic faces [14,44] while others have found more accuracy in the recognition of anger when expressed by human faces [45].

As regards as the age of stimuli, was observed that young faces were better decoded than older ones in the case of anger and surprise, middle-aged faces were better decoded when expressing sadness, fear and happiness compared to young and old faces, while in the case of disgust, old faces where better decoded compared to young ones. Conclusively, as regard as the gender of stimuli, emerged that female faces where better decoded compared to male ones in the case of disgust, anger, sadness, fear, and happiness, while neutral facial expressions where better decoded when conveyed by male rather than females. Once again, the effects of variables such as the age and the gender of stimuli seem to not have a direct and univocal effect on emotion recognition abilities, somewhat appears that this effect, as that of other variables previously listed, could be mediated by the emotional category administered.

Was also observed a high percentage of 'other emotion' selection especially concerning naturalistic stimuli, this result could be linked to the fact that giving this option to participants without advising them that selecting “other emotion” would have resulted in an error, may have made participants think that it was a valid response.

Conclusively, we observed that the emotional categories better decoded were Happiness and Anger, followed by Neutrality, Sadness and Surprise. Disgust and Fear were the emotional categories worse decoded by participants.

## VI. LIMITATIONS

It is necessary to highlight those inferences born from the obtained results are affected and limited by design of the experiment and the followed experimental procedures. A possible limitation of the proposed investigation consists in the fact that we do not consider the change in position of the showed faces, a variable that in other studies was found significantly affecting emotion recognition rates [16]; it is likely that this could have affected results of the naturalistic experiment, for this reason future studies should consider this issue.

It is also necessary to test the effects of other typology of stimuli as for instance dynamic and multimodal facial expressions and not only static facial expressions, since as already highlighted [46] the interaction with intelligent systems is actually continuous and multimodal, and emotions conveyed by more dynamic virtual agents are better decoded [47]. Indeed, we observed recognition percentages below the 50% threshold for some emotions as disgust, fear and surprise, both within the synthetic and naturalistic experiments. This issue could be linked to the fact that the stimuli are all static images, while certain emotions need more information to be accurately recognized, such as those coming from dynamic facial expressions.

Moreover, it is necessary to highlight that participants' number and gender is not always perfectly balanced among the groups, however, it is our opinion that differences related to participants' number and gender are too modest to affect results.

Another possible limitation is related to the fact that naturalistic faces were taken from popular movies, this could have facilitated participants in the recognition task, as they could have remembered the context which surround the emotional expression. However, we believe that if participants' knowledge of the context which surround the emotional faces would have affected results, we should have found a facilitation effect, or in other words, we should have found that naturalistic task's scores were systematically better than the synthetic ones, but this is not the case as showed by our results.

A further limitation could be represented by the choice of having administered the synthetic and the naturalistic experiments to different groups of participants, potentially limiting the possibility of comparing the two groups significantly. However, we have to highlight that we could not have the same subjects doing both experiments even if at a distance of time to avoid a learning effect and therefore to facilitate the test, or in any case to avoid any effect of the first task on the second one. However, we have tried to form as homogeneous groups as possible to avoid not meaningful comparisons between them.

## VII. CONCLUSIONS

An investigation was proposed introducing a study conducted with the aim of exploring how differently aged (young, middle-aged and elders) groups of participants decode static facial emotional expressions conveyed by differently aged (young, middle-aged and elders) human beings and virtual agents, moreover, gender effects of both participants and the proposed emotional faces were also considered as study variables to be investigated. Understanding how these factors affect the way people decode emotions expressed by virtual agents, could be helpful in order to improve human-agents interaction. Participants, depending on the group they belong to, were required to complete a naturalistic or a synthetic emotion recognition task (depending on whether the face to be decoded belonged to a human being or to a virtual agent) attaching an emotional label to each static picture choosing among disgust, anger, fear, sadness, happiness, surprise, and neutrality.

The reason behind the choice of this experimental design, in which we did not administered participants both experiments, is due to the high number of stimuli and the duration of the experiments; our aim was to avoid participants' fatigue, which could have affected their performances. We observed that young participants showed better recognition performances of anger, sadness, and neutrality. Female participants showed better recognition performances of sadness, fear, and neutrality; sadness and fear were better recognized when conveyed by naturalistic faces, while happiness, surprise and neutrality were better recognized when represented by synthetic faces. Young facial expressions were better decoded when conveying anger and surprise, middle-aged ones were better decoded when expressing sadness, fear, and happiness while old facial expressions were better decoded when conveying disgust; concerning the gender of the showed faces, on average female ones where better decoded compared to males.

What differentiates our study from those conducted so far is that, for instance, several researches have tested differences in the recognition of natural and synthetic emotions using the same face while conveying different emotional expressions, or in some cases participants were shown faces belonging to only one gender, not allowing to take into account the influence that the gender of the face which expresses an emotion can exert on emotion recognition processes, as well as the fact that showing always the same face to participants could result in a task facilitation effect. Other studies are based on experiments in which participants were even shown faces of non-humanoid agents, namely agents with animal-like resemblance. Our work, in comparison with previous approaches, consisted in involving differently aged groups of participants comparing their performances, in order to understand how the ability to decode synthetic and natural emotional expressions is affected by aging. Moreover, in our experiments were developed and exploited virtual agents with variegate characteristics and of different gender. The novelty of our work, in particular, is represented by the fact that were developed and used differently aged virtual agents (young, middle-aged, and old) and agents with truly variegated physical features, allowing to develop a full-fledged dataset of emotional virtual agents.

**APPENDIX A**:

Detailed description of the statistical analyses performed in the present study. The reason why we used repeated measures ANOVA is because this kind of test allows to test both the effects of within and between subjects’ variables. More specifically, to evaluate whether differences among recognition scores attributed by the differently aged groups and groups who saw different type of stimuli (naturalistic and synthetic) are statistically significant and the considered variables related to the showed faces (age and gender) significantly affected results. Repeated measures ANOVA compare marginal means for each stimuli variables (the within factors) and each differently gendered and aged population and type of stimuli administered (the between factors). Results are expressed in terms of F and p values, where F (called Fisher value) is a ratio of two variances and p provides the significance cut-off (set to .05). When testing a set of different variables for statistical significance across various groups, some of the variables may be falsely considered as statistically significant. The purpose of Bonferroni’s post-hoc tests (a multiple testing correction) is to keep the overall error rate/false positives to less than the specified p-value cut-off. ANOVA analyses were chosen because of the experimental set-up requiring the involvement of three differently aged groups in both of the synthetic and naturalistic experiments. The age and gender of participants and type of stimulus administered were considered between factors and age and gender of stimuli as within factors.

**A.1: Statistical Analyses performed for Section IV:**
**Effects of participants age and type of stimulus on decoding accuracy of emotional expressions.**

A first elaboration of the acquired data was performed with the aim to assess participants’ ability to correctly decode the proposed emotional categories (i.e., disgust, anger, sadness, fear, happiness, surprise, and neutrality) independently from the age and the gender of the faces. Repeated measures ANOVA were performed on the collected data, considering participants’ gender, age groups (young, middle-aged, elders) and type of stimulus administered (naturalistic vs synthetic facial expressions) as between subjects’ variables, and total decoding scores of the proposed emotional category as within subjects’ variables. The significance level was set at α <.05 and differences among means were assessed through Bonferroni’s post hoc tests.**Significant effects of participants’ age emerged** (F (2,266) = 21.247, p<<.01). Bonferroni post hoc tests revealed that this was due to young participants (mean=3.050, SD=.056) who showed better decoding performances compared to middle-aged (mean=2.639, SD=.056, p<<.01) and old participants (mean=2.570, SD=.057, p<<.01). **Significant effects of the type of stimulus emerged** (F (1,266) = 11.871, p=.001). Bonferroni post hoc tests revealed that this was due to synthetic stimuli (mean=2.865, SD=.046) which were better decoded compared to naturalistic ones (mean=2.641, SD=.047, p=.001).

**Significant differences** (F (6, 1596) =216.956, p<<.01) **emerged as regards as participants’ recognition accuracy of the proposed emotional category.** Each emotional category significantly differed from each other: Fear (mean=1.157, SD=.062), Disgust (mean=1.613, SD=.077), Surprise (mean=2.469, SD=.087), Sadness (mean=2.725, SD=.073), Neutrality (mean=3.046, SD=.095), Happiness (mean=3.938, SD=.081), Anger (mean=4.324, SD=.075), p<<.01; between anger and sadness the difference was slightly lower (p=.001). The only exception was represented by Sadness, which did not significantly differ from Surprise (p=.620) and Neutrality (p=.138). **A significant interaction was observed** (F (2, 266) = 4.037, p=.019) **between participants’ age and the type of stimulus administered.** Bonferroni’s post hoc tests were performed for each single factor (participants’ age and stimulus type). These tests revealed that:

a)Concerning participants’ age: among the groups who were administered naturalistic stimuli, young participants (mean=2.916, SD=.081) showed better recognition performances compared to middle-aged participants (mean=2.425, SD=.081, p<<.01) and old participants (mean=2.581, SD=.081, p=.011), the same occurred among the groups who were administered synthetic facial expressions, with young participants (mean=3.185, SD=.078) showing better performances compared to middle-aged (mean=2.853, SD=.078, p=.009) and old participants (mean=2.559, SD=.081, p=.029).

b)Concerning the stimulus type: young participants who were administered synthetic stimuli (mean=3.185, SD=.078) showed better performances compared to the group of young participants who were administered naturalistic stimuli (mean=2.916, SD=.081, p=.017); the same happened for the group of middle-aged participants who were administered synthetic stimuli (mean=2.853, SD=.078) which showed better performances compared to the group of middle-aged participants who were administered naturalistic stimuli (mean=2.425, SD=.081, p<<.01).

**A significant interaction emerged** (F (6, 1596) =4.691, p<<.01) **between participants’ gender and the emotional categories considered.** Bonferroni’s post hoc tests were performed for each single factor (participants’ gender and emotional category). These tests revealed that:

a) Concerning participants' gender: sadness was better decoded by female participants (mean=2.975, SD=.099) compared to male participants (mean=2.474, SD=.108, p=.001); fear was better decoded by female participants (mean=1.302, SD=.084) compared to male participants (mean=1.012, SD=.091, p=.019) and neutrality was better decoded by male participants (mean=3.327, SD=.140) compared to female participants (mean=2.408, SD=.118, p=.003).

b) Concerning emotional category: Male participants showed differences in the recognition of each emotional category: Fear (mean=1.012, SD=.091), Disgust (mean=1.574, SD=.113), Surprise (mean=2.530, SD=.128), Sadness (mean=2.474, SD=.108), Neutrality (mean=3.327, SD=.140), Happiness (mean=3.941, SD=.119), Anger (mean=4.377, SD=.111), with p<<.01; lower differences were found between happiness and Anger (p=.027) and happiness and Neutrality (p=.009), as well as between disgust and fear (p=.002). The only exception was represented by Sadness and Surprise which did not significantly differ from each other (p=.1000). Female participants showed differences in the recognition of each emotional category: Fear (mean=1.302, SD=.084), Disgust (mean=1.652, SD=.104), Surprise (mean=2.408, SD=.118), Sadness (mean=2.975, SD=.099), Neutrality (mean=2.766, SD=.128), Happiness (mean=3.935, SD=.109), Anger (mean=4.271, SD=.102), with p<<.01; lower differences were found between Sadness and Surprise (p=.008). The only exception was represented by Disgust and Fear (p=.173), Anger and Happiness (p=.147), Sadness and Neutrality (p=.1000), Surprise and Neutrality (p=.592) which did not significantly differed from each other.

**A significant interaction emerged** (F (12, 1596) =2.525, p=.003) **between participants' age and the emotional categories considered.** Bonferroni's post hoc tests were performed for each single factor (participants' age and emotional category). These tests revealed that:

a) Concerning participants' age: anger was better decoded by young participants (mean=4.669, SD=.129) compared to middle-aged (mean=4.187, SD=.130, p=.027) and old participants (mean=4.116, SD=.132, p=.009); sadness was better decoded by young (mean=3.040, SD=.126) rather than old participants (mean=2.459, SD=.128, p=.004); surprise was worse decoded by old participants (mean=1.936, SD=.152), compared to middle-aged (mean=2.668, SD=.150, p=.002) and young participants (mean=2.803, SD=.149, p<<.01); neutrality was better decoded by young participants (mean=3.606, SD=.163) compared to middle-aged (mean=2.712, SD=.164, p<<.01) and old participants (mean=2.821, SD=.166, p=.003).

b) Concerning emotional category: Young participants showed differences in the recognition of each emotional category: Fear (mean=1.269, SD=.106), Disgust (mean=1.881, SD=.132), Surprise (mean=2.803, SD=.149), Sadness (mean=3.040, SD=.126), Neutrality (mean=3.606, SD=.163), Happiness (mean=4.083, SD=.139), Anger (mean=4.669, SD=.129), with p<<.01; lower differences were found between Disgust and Fear (p=.006), between happiness and Anger (p=.005), and between Surprise and Neutrality (p=.002). The only exception was represented by Neutrality compared to Sadness (p=.114) and Happiness (p=.378), and Sadness compared to Surprise (p=.1000) which did not significantly differ from each other. Middle-aged participants showed differences in the recognition of each emotional category: Fear (mean=1.050, SD=.107), Disgust (mean=1.499, SD=.133), Surprise (mean=2.668, SD=.150), Sadness (mean=2.675, SD=.127), Neutrality (mean=2.712, SD=.164), Happiness (mean=3.682, SD=.140), Anger (mean=4.187, SD=.130), with p<<.01; lower differences were found between happiness and Anger (p=.032). The only exception was represented by Disgust and Fear (p=.163), Sadness and Surprise (p=.1000), Sadness and Neutrality (p=.1000), Surprise and Neutrality (p=.1000) which did not significantly differed from each other. Old participants showed differences in the recognition of each emotional category: Fear (mean=1.152, SD=.108), Disgust (mean=1.459, SD=.135), Surprise (mean=1.936, SD=.152), Sadness (mean=2.459, SD=.128), Neutrality (mean=2.821, SD=.166), Happiness (mean=4.049, SD=.142), Anger (mean=4.116, SD=.132), with p<<.01; lower differences were found between surprise and fear (p=.003) and between surprise and neutrality (p=.001). The only exception was represented by Disgust and Fear (p=.100), Disgust and Surprise (p=.366), Anger and Happiness (p=.1000), Sadness and Surprise (p=.234), Sadness and neutrality (p=.1000) which did not significantly differed from each other.

**A significant interaction emerged** (F (6, 1596) =19.388, p<<.01) **between the type of stimulus administered (naturalistic vs synthetic) and the emotional categories considered**. Bonferroni's post hoc tests were performed for each single factor (stimulus type and emotional category). These tests revealed that:

a)Concerning stimulus type: sadness was better decoded by groups who were administered naturalistic stimuli (mean=3.035, SD=.105) rather than synthetic ones (mean=2.414, SD=.103 p<<.01); fear was better decoded by groups who were administered naturalistic stimuli (mean=1.435, SD=.088) rather than synthetic ones (mean=.879, SD=.086, p<<.01); happiness was better decoded by groups who were administered synthetic stimuli (mean=4.228, SD=.113) rather than naturalistic ones (mean=3.649, SD=.116, p<<.01); surprise was better decoded by groups who were administered synthetic stimuli (mean=2.777, SD=.121) rather than naturalistic ones (mean=2.161, SD=.124, p<<.01); neutrality was better decoded by groups who were administered synthetic stimuli (mean=3.704, SD=.133) rather than naturalistic ones (mean=2.388, SD=.135, p<<.01).

b) Concerning emotional category: significant differences were observed in the recognition Naturalistic faces of Fear (mean=1.435, SD=.088), Disgust (mean=1.549, SD=.110), Surprise (mean=2.161, SD=.124), Sadness (mean=3.035, SD=.105), Neutrality (mean=2.388, SD=.135), Happiness (mean=3.649, SD=.116), Anger (mean=4.269, SD=.107), with p<<.01; lower differences were found between Disgust and Surprise (p=.004), between sadness when compared whit happiness (p=.002) and neutrality (p=.003). The only exception was represented by disgust compared to fear (p=.1000) and Surprise compared to neutrality (p=.1000) which did not significantly differ from each other. Significant differences were also observed in the recognition Synthetic faces of Fear (mean=.879, SD=.086), Disgust (mean=1.678, SD=.108), Surprise (mean=2.777, SD=.121), Sadness (mean=2.414, SD=.103), Neutrality (mean=3.704, SD=.133), Happiness (mean=4.228, SD=.113), Anger (mean=4.379, SD=.105), with p<<.01; lower differences were found between Anger and Neutrality (p=.001), between happiness when compared whit neutrality (p=.032). The only exception was represented by anger compared to happiness (p=.1000), and Sadness compared to surprise (p=.568) which did not significantly differ from each other.

## A.2: Statistical Analyses performed for Section IV: Effects of participants' and stimuli's age and gender on the decoding accuracy of naturalistic and synthetic emotional faces.

To test the effect of some variables as the age and the gender of the showed faces on emotion decoding accuracy, a further ANOVA repeated measures analyses were carried out for each emotional category (disgust, anger, fear, sadness, happiness, surprise, and neutrality) considering participants' age group (young, middle-aged, and elders), gender and the type of stimulus administered (naturalistic vs synthetic facial expressions) as between subjects' factors, and age and gender of stimuli as within factors. The significance level was set at α <.05 and differences among means were assessed through Bonferroni's post hoc tests.

*Disgust*

**Significant effects of participants' age emerged** (F (2,266) = 3.093, p=.047). Bonferroni's post hoc multiple adjustments do not allow to identify which means produced these differences.

**Significant effects of the age of stimuli were observed** (F (2, 532) = 4.691, p=.010). Bonferroni post hoc tests revealed that this was due to old facial expressions (mean=.297, SD=.019) which were better recognized compared to young aged facial expressions (mean=.229, SD=.017, p=.011). **Significant effects of the gender of stimuli were observed** (F (1, 266) = 24.680, p<<.01). Bonferroni post hoc tests revealed that this was due to female facial expressions (mean=.315, SD=.017) which were better recognized compared to male facial expressions (mean=.223, SD=.014, p<<.01).

**A significant interaction emerged** (F (2,266) = 5.069, p=.007) **between participants' age and the type of stimulus (naturalistic vs synthetic) administered.** Bonferroni's post hoc tests were performed for each single factor (participants' age and type of stimuli). These tests revealed that:

a) Concerning participants' age: the group of young participants decoding naturalistic faces showed better decoding performances (mean=.348, SD=.032) compared to the group of middle-aged (mean=.186, SD=.032, p=.001) participants.

b) Concerning stimuli's type: the group of middle-aged participants required to decode synthetic facial expressions showed better performances (mean=.314, SD=.031) compared to the group of middle-aged required to decode naturalistic facial expressions (mean=.186, SD=.032 p=.004).

**A significant interaction emerged** (F (2, 532) = 32.124, p<<.01) **between stimuli's age and gender**. Bonferroni's post hoc tests were performed for each single factor (age and gender of stimuli). These tests revealed that:

a) Concerning stimuli's age: when decoding female facial expressions, participants better recognized disgust conveyed by middle-aged faces (mean=.399, SD=.029) compared to young (mean=.169, SD=.022, p<<.01) and old faces (mean=.377, SD=.028, p<<.01). When decoding males, participants better decoded young faces (mean=.290, SD=.023) compared to middle-aged ones (mean=.162, SD=.022, p<<.01).

b) Concerning stimuli's gender: when decoding young faces participants better decoded males (mean=.290, SD=.023) rather than female faces (mean=.169 , SD=.022, p<<.01); when decoding middle-aged and old faces they better recognized female faces (middle-aged female faces=.399, SD=.029; old female faces=.377, SD=.028) compared to male ones (middle-aged male faces=.162, SD=.022; old male faces=.217, SD=.023, p<<.01).

**A significant interaction emerged** (F (2, 532) = 7.341, p=.001) **between stimuli's age and the type of stimulus (naturalistic vs synthetic) administered.** Bonferroni's post hoc tests were performed for each single factor (type and age of stimuli). These tests revealed that:

a) Concerning the type of stimulus: young facial expressions were better decoded by the groups required to decode naturalistic facial expressions (mean=.269, SD=.025) compared to the groups required to label synthetic facial expressions (mean=.190, SD=.024, p=.023), while middle-aged faces were better recognized by the groups administered with synthetic facial expressions (mean=.325, SD=.027) compared to the groups required to recognize naturalistic faces (mean=.236, SD=.027, p=.020).

b) Concerning age of stimuli: the groups required to decode synthetic facial expressions worse decoded young facial expressions (mean=.190, SD=.024) compared to middle-aged (mean=.325, SD=.027, p<<.01) and old ones (mean=.324, SD=.027, p<<.01).

**A significant interaction emerged** (F (2, 266) = 3.625, p=.028) **between stimuli's gender and participants' age.** Bonferroni's post hoc tests were performed for each single factor (participants' age and gender of stimuli). These tests revealed that:

a) Concerning participants' age: female faces where better recognized by young participants (mean=.370, SD=.029) compared to old participants (mean=.255, SD=.030, p=.019).

b) Concerning the gender of stimuli: young participants better decoded females (mean=.370, SD=.029) rather than male faces (mean=.257, SD=.025, p<<.01), and the same happened for middle-aged participants who better decoded females (mean=.320, SD=.030) rather than male faces (mean=.180, SD=.025, p<<.01).

*Anger*

**Significant effects of participants' age emerged** (F (2,266) = 5.378, p=.005). Bonferroni post hoc tests revealed that this was due to young participants (mean=.778, SD=.022) who showed better decoding performances compared to middle-aged (mean=.698, SD=.022, p=.027) and old participants (mean=.686, SD=.022, p=.009). **Significant effects of the age of stimuli were observed** (F (2, 532) = 31.610, p<<.01). Bonferroni post hoc tests revealed that this was due to young facial expressions (mean=.816, SD=.017) which were better recognized compared to middle-aged (mean=.656, SD=.014, p<<.01) and old facial expressions (mean=.691, SD=.021, p<<.01). **Significant effects of the gender of stimuli were observed** (F (1, 266) = 6.828, p=.009). Bonferroni post hoc tests revealed that this was due to female facial expressions (mean=.746, SD=.017) which were better recognized compared to male facial expressions (mean=.696, SD=.015, p=.009). **A significant interaction emerged** (F (2, 532) = 122.539, p<<.01) **between stimuli's age and gender.** Bonferroni's post hoc tests were performed for each single factor (age and gender of stimuli). These tests revealed that:

a) Concerning stimuli's age: when decoding female facial expressions, participants better recognized anger conveyed by middle-aged (mean=.864, SD=.020) compared to young (mean=.754, SD=.024, p<<.01) and old faces (mean=.618, SD=.029, p<<.01). When decoding males, participants better decoded young faces (mean=.877, SD=.019) compared to middle-aged (mean=.447, SD=.019, p<<.01) and old ones (mean=763, SD=.026, p<<.01).

b) Concerning stimuli's gender: when decoding young faces participants better decoded males (mean=.877, SD=.019) rather than female faces (mean=.754, SD=.024, p<<.01); when decoding middle-aged faces, they better recognized female (mean=.864, SD=.020) rather than males (mean=.447, SD=.019, p<<.01); when decoding old faces participants better decoded males (mean=.763, SD=.026) rather than female faces (mean=.618, SD=.029, p<<.01).

**A significant interaction emerged** (F (2, 532) = 94.543, p<<.01) **between stimuli's age and the type of stimulus (naturalistic vs synthetic) administered.** Bonferroni's post hoc tests were performed for each single factor (type and age of stimuli). These tests revealed that:

a) Concerning the type of stimulus: young facial expressions were better decoded by the groups required to decode naturalistic facial expressions (mean=.934, SD=.024) compared to the groups required to label synthetic facial expressions (mean=.698, SD=.024, p<<.01), while middle-aged faces were better recognized by the groups administered with synthetic facial expressions (mean=.824, SD=.020) compared to the groups required to recognize naturalistic faces (mean=.487, SD=.020, p<<.01).

b) Concerning age of stimuli: the groups required to decode naturalistic facial expressions better decoded young facial expressions (mean=.934, SD=.024) compared to middle-aged (mean=.487, SD=.020, p<<.01) and old ones (mean=.714, SD=.030, p<<.01). The groups required to decode synthetic facial expressions better decoded middle-aged facial expressions (mean=.824, SD=.020) compared to young (mean=.698, p<<.01) and old ones (mean=.668, SD=.029, p<<.01).

**A significant interaction emerged** (F (1, 266) = 126.915, p<<.01) **between stimuli's gender and the type of stimulus (naturalistic vs synthetic) administered.** Bonferroni's post hoc tests were performed for each single factor (type and gender of stimuli). These tests revealed that:

a) Concerning the type of stimulus: female facial expressions were better decoded by the groups required to decode naturalistic facial expressions (mean=.844, SD=.024) compared to the groups required to label synthetic facial expressions (mean=.647, SD=.024, p<<.01), while male faces were better recognized by the groups administered with synthetic facial expressions (mean=.812, SD=.020) compared to the groups required to recognize naturalistic faces (mean=.579, SD=.021, p<<.01).

b) Concerning gender of stimuli: the groups required to decode naturalistic facial expressions better decoded female facial expressions (mean=.844, SD=.024) compared to male ones (mean=.579, SD=.021, p<<.01). The groups required to decode synthetic facial expressions better decoded male facial expressions (mean=.812, SD=.020) compared to female ones (mean=.647, SD=.024, p<<.01).

*Sadness*

**Significant effects of participants' gender emerged** (F (1,266) = 11.683, p=.001). Bonferroni post hoc tests revealed that this was due to female participants (mean=.496, SD=.017) who showed better decoding performances compared to male participants (mean=.412, SD=.018, p=.001). **Significant effects of participants' age emerged** (F (2,266) = 5.352, p=.005). Bonferroni post hoc tests revealed that this was due to young participants (mean=.507, SD=.021) who showed better decoding performances compared to old participants (mean=.410, SD=.021, p=.004). **Significant effects of the type of stimulus emerged** (F (1,266) = 17.955, p<<.01). Bonferroni post hoc tests revealed that this was due to naturalistic stimuli (mean=.506, SD=.017) were better decoded compared to synthetic ones (mean=.402, SD=.017, p<<.01). **Significant effects of the age of stimuli were observed** (F (2, 532) = 6.429, p=.002). Bonferroni post hoc tests revealed that this was due to middle-aged facial expressions (mean=.502, SD=.019) which were better recognized compared to young (mean=.443, SD=.018, p=.046) and old facial expressions (mean=.417, SD=.019, p=.001). **Significant effects of the gender of stimuli were observed** (F (1, 266) = 9.597, p=.002). Bonferroni post hoc tests revealed that this was due to female facial expressions (mean=.487, SD=.015) which were better recognized compared to male facial expressions (mean=.421, SD=.017, p=.002). **A significant interaction emerged** (F (2, 532) = 50.987, p<<.01) **between stimuli's age and gender.** Bonferroni's post hoc tests were performed for each single factor (age and gender of stimuli). These tests revealed that:

a) Concerning stimuli's age: when decoding female facial expressions, participants better recognized sadness conveyed by middle-aged (mean=.654, SD=.024) compared to young (mean=.465, SD=.023, p<<.01) and old faces (mean=.341, SD=.024, p<<.01). When decoding males, participants better decoded old faces (mean=.494, SD=.030) compared to middle-aged (mean=.350, SD=.024, p<<.01) and young ones (mean=420, SD=.025, p<<.01).

b) Concerning stimuli's gender: when decoding middle-aged faces participants better recognized female (mean=.654, SD=.024) rather than males (mean=.350, SD=.024, p<<.01); when decoding old faces participants better decoded males (mean=.494, SD=.030) rather than female faces (mean=.341, SD=.024, p<<.01).

**A significant interaction emerged** (F (2, 532) = 149.160, p<<.01) **between stimuli's age and the type of stimulus (naturalistic vs synthetic) administered.** Bonferroni's post hoc tests were performed for each single factor (type and age of stimuli). These tests revealed that:

a) Concerning the type of stimulus: young facial expressions were better decoded by the groups required to decode naturalistic facial expressions (mean=.727, SD=.025) compared to the groups required to label synthetic facial expressions (mean=.158, SD=.025, p<<.01), while old faces were better recognized by the groups administered with synthetic facial expressions (mean=.541, SD=.026) compared to the groups required to recognize naturalistic faces (mean=.294, SD=.027, p<<.01).
b) Concerning age of stimuli: the groups required to decode naturalistic facial expressions better decoded young facial expressions (mean=.727, SD=.025) compared to middle-aged (mean=.497, SD=.028, p<<.01) and old ones (mean=.294, SD=.027, p<<.01). The groups required to decode synthetic facial expressions worse decoded young facial expressions (mean=.158, SD=.025) compared to middle-aged (mean=.508, SD=.027, p<<.01) and old ones (mean=.541, SD=.026, p<<.01).
**A significant interaction emerged** (F (1, 266) = 22.968, p<<.01) **between stimuli's gender and the type of stimulus (naturalistic vs synthetic) administered.** Bonferroni's post hoc tests were performed for each single factor (type and gender of stimuli). These tests revealed that:
a) Concerning the type of stimulus: female facial expressions were better decoded by the groups required to decode naturalistic facial expressions (mean=.589, SD=.021) compared to the groups required to label synthetic facial expressions (mean=.384, SD=.021, p<<.01).
b) Concerning gender of stimuli: the groups required to decode naturalistic facial expressions better decoded female facial expressions (mean=.589, SD=.021) compared to male ones (mean=.423, SD=.025, p<<.01).

*Fear*
**Significant effects of participants' gender emerged** (F (1,266) = 5.526, p=.019). Bonferroni post hoc tests revealed that this was due to female participants (mean=.217, SD=.014) who showed better decoding performances compared to male participants (mean=.169, SD=.015, p=.019). **Significant effects of the type of stimulus emerged** (F (1,266) = 20.290, p<<.01). Bonferroni post hoc tests revealed that this was due to naturalistic stimuli (mean=.239, SD=.015) were better decoded compared to synthetic ones (mean=.146, SD=.014, p<<.01). **Significant effects of the age of stimuli were observed** (F (2, 532) = 37.582, p<<.01). Bonferroni post hoc tests revealed that this was due to middle-aged facial expressions (mean=.289, SD=.017) which were better recognized compared to young (mean=.132, SD=.013, p<<.01) and old facial expressions (mean=.157, SD=.016, p<<.01). **Significant effects of the gender of stimuli were observed** (F (1, 266) = 156.252, p<<.01). Bonferroni post hoc tests revealed that this was due to female facial expressions (mean=.301, SD=.016) which were better recognized compared to male facial expressions (mean=.084, SD=.011, p<<.01). **A significant interaction emerged** (F (2, 532) = 70.503, p<<.01) **between stimuli's age and gender.** Bonferroni's post hoc tests were performed for each single factor (age and gender of stimuli). These tests revealed that:
a) Concerning stimuli's age: when decoding female facial expressions, participants better recognized fear conveyed by middle-aged (mean=.525, SD=.030) compared to young (mean=.228, SD=.023, p<<.01) and old faces (mean=.151, SD=.022, p<<.01). When decoding males, participants better decoded old faces (mean=.163, SD=.022) compared to middle-aged (mean=.053, SD=.013, p<<.01) and young ones (mean=.037, SD=.012, p<<.01).
b) Concerning stimuli's gender: when decoding young faces participants better recognized female (mean=.228, SD=.023) rather than males (mean=.037, SD=.012, p<<.01) and the same occurred when decoding middle-aged faces, as participants better decoded females (mean=.525, SD=.030) rather than male faces (mean=.053, SD=.013, p<<.01).
**A significant interaction emerged** (F (2, 532) = 3.447, p=.033) **between stimuli's age and the type of stimulus (naturalistic vs synthetic) administered.** Bonferroni's post hoc tests were performed for each single factor (type and age of stimuli). These tests revealed that:
a) Concerning the type of stimulus: young facial expressions were better decoded by the groups required to decode naturalistic facial expressions (mean=.208, SD=.019) compared to the groups required to label synthetic facial expressions (mean=.057, SD=.018, p<<.01); old faces were better recognized by the groups administered with naturalistic facial expressions (mean=.189, SD=.022) compared to the groups required to recognize synthetic faces (mean=.125, SD=.022, p=.045).
b) Concerning age of stimuli: the groups required to decode naturalistic facial expressions better decoded middle-aged facial expressions (mean=.321, SD=.024) compared to young (mean=.208, SD=.019, p<<.01) and old ones (mean=.189, SD=.022, p<<.01). The groups required to decode synthetic facial expressions better decoded middle-aged facial expressions (mean=.258, SD=.023) compared to young (mean=.057, SD=.018, p<<.01) and old ones (mean=.125, SD=.022, p<<.01).
**A significant interaction was observed** (F (1, 266) = 3.986, p=.047) **between gender of stimuli and participants' gender.** Bonferroni's post hoc tests were performed for each single factor (participants' and stimuli's gender). These tests revealed that:
a) Concerning participants' gender: females' facial expressions were better decoded by female participants (mean=.343, SD=.021) compared to male participants (mean=.260, SD=.023, p=.009).
b) Concerning gender of stimuli: male participants better decoded females' facial expressions (mean=.260, SD=.023) compared to male ones (mean=.077, SD=.016, p<<.001); female participants better decoded female facial expressions (mean=.343, SD=.021) compared to male ones (mean=.091, SD=.014, p<<.01).
**A significant interaction was observed** (F (2, 266) = 3.463, p=.033) **between gender of stimuli and participants' age**. Bonferroni's post hoc tests were performed for each single factor (participants' age and gender of stimuli). These tests revealed that:
a) Concerning participants' age, no significant differences emerged.
b) Concerning gender of stimuli: young participants better recognized female faces (mean=.338, SD=.027) compared to male ones (mean=.085, SD=.018, p<<.01), and same occurred for middle-aged participants who better recognized female faces (mean=.251, SD=.027) compared to male ones (mean=.099, SD=.018, p<<.01), and for old participants who better recognized female faces (mean=.315, SD=.028) compared to male ones (mean=.069, SD=.019, p<<.01).

*Happiness*

**Significant effects of the type of stimulus emerged** (F (1,266) = 12.818, p<<.01). Bonferroni post hoc tests revealed that this was due to synthetic stimuli (mean=.705, SD=.019) which were better decoded compared to naturalistic ones (mean=.608, SD=.019, p<<.01). **Significant effects of the age of stimuli were observed** (F (2, 532) = 127.488, p<<.01). Bonferroni post hoc tests revealed that this was due to middle-aged facial expressions (mean=.856, SD=.016) which were better recognized compared to young (mean=.600, p<<.01) and old facial expressions (mean=.513, SD=.019, p<<.01). **Significant effects of the gender of stimuli were observed** (F (1, 266) = 302.636, p<<.01). Bonferroni post hoc tests revealed that this was due to female facial expressions (mean=.816, SD=.016) which were better recognized compared to male facial expressions (mean=.496, SD=.016, p<<.01). **A significant interaction emerged** (F (2, 532) = 53.378, p<<.01) **between stimuli's age and gender.** Bonferroni's post hoc tests were performed for each single factor (age and gender of stimuli). These tests revealed that:

a) Concerning stimuli's age: when decoding female facial expressions, participants better recognized happiness conveyed by middle-aged (mean=.894, SD=.019) compared to young (mean=.781, SD=.025, p<<.01) and old faces (mean=.773, SD=.026, p<<.01). When decoding males, participants better decoded middle-aged faces (mean=.818, SD=.023) compared to young (mean=.418, SD=.029, p<<.01) and old ones (mean=.254, SD=.024, p<<.01).

b) Concerning stimuli's gender: when decoding young faces participants better recognized female (mean=.781, SD=.025) rather than males (mean=.418, SD=.029, p<<.01), the same occurred when decoding middle-aged faces, as participants better decoded females (mean=.894, SD=.019) rather than male faces (mean=.818, SD=.023, p=.003) and when decoding old faces, as participants better decoded females (mean=.773, SD=.026) rather than male faces (mean=.254, SD=.024, p<<.01).

**A significant interaction was observed** (F (4, 532) = 4.834, p=.001) **between age of stimuli and participants' age.** Bonferroni's post hoc tests were performed for each single factor (participants' age and age of stimuli). These tests revealed that:

a) Concerning participants' age: young faces were better decoded by young participants (mean=.684, SD=.035) compared to middle-aged participants (mean=.502, SD=.035, p=.001).

b) Concerning age of stimuli: young participants better decoded middle-aged facial expressions (mean=.819, SD=.028) compared to young (mean=.684, SD=.035, p=.001) and old faces (mean=.539, SD=.033, p<<.01); middle-aged participants better decoded middle-aged facial expressions (mean=.838, SD=.028) compared to young (mean=.502, SD=.035, p<<.01) and old faces (mean=.501, SD=.033, p<<.01); old participants better decoded middle-aged facial expressions (mean=.912, SD=.029) compared to young (mean=.613, SD=.035, p<<.01) and old faces (mean=.500, SD=.033, p<<.01).

**A significant interaction emerged** (F (2, 532) = 9.525, p=<<.01) **between stimuli's age and the type of stimulus (naturalistic vs synthetic) administered.** Bonferroni's post hoc tests were performed for each single factor (type and age of stimuli). These tests revealed that:

a) Concerning the type of stimulus: old faces were better recognized by the groups administered with synthetic facial expressions (mean=.618, SD=.027) compared to the groups required to recognize naturalistic faces (mean=.409, SD=.027, p<<.01).

b) Concerning age of stimuli: the groups required to decode naturalistic facial expressions better decoded middle-aged facial expressions (mean=.840, SD=.024) compared to young (mean=.575, SD=.029, p<<.01) and old ones (mean=.409, SD=.027, p<<.01). The groups required to decode synthetic facial expressions better decoded middle-aged facial expressions (mean=.872, SD=.023) compared to young (mean=.62, SD=. 028, p<<.01) and old ones (mean=.618, SD=.027, p<<.01).

**A significant interaction emerged** (F (1, 266) = 70.549, p<<.01) **between stimuli's gender and the type of stimulus (naturalistic vs synthetic) administered.** Bonferroni's post hoc tests were performed for each single factor (type and gender of stimuli). These tests revealed that:

a) Concerning the type of stimulus: male facial expressions were better decoded by the groups required to decode synthetic facial expressions (mean=.622, SD=.023) compared to the groups required to label naturalistic facial expressions (mean=.371, SD=.024, p<<.01).

b) Concerning gender of stimuli: the groups required to decode naturalistic facial expressions better decoded female facial expressions (mean=.845, SD=.023) compared to male ones (mean=.371, SD=.024, p<<.01) and the same occurred for the groups required to decode synthetic facial expressions which better decoded female facial expressions (mean=.787, SD=.023) compared to male ones (mean=.622, SD=.023, p<<.01).

*Surprise*

**Significant effects of participants' age emerged** (F (2,266) = 9.538, p<<.01). Bonferroni post hoc tests revealed that this was due to old participants (mean=.323, SD=.025) who showed worse decoding performances compared to middle-aged (mean=.445, SD=.025, p=.002) and young participants (mean=.467, SD=.025, p<<.01). **Significant effects of the type of stimulus emerged** (F (1,266) = 12.260 p<<.01). Bonferroni post hoc tests revealed that this was due to synthetic stimuli (mean=.463, SD=.020) which were better decoded compared to naturalistic ones (mean=.360, SD=.021, p<<.01).

**Significant effects of the age of stimuli were observed** (F (2, 532) = 16.489, p<<.01). Bonferroni post hoc tests revealed that this was due to young facial expressions (mean=.499, SD=.023) which were better recognized compared to middle-aged (mean=.354, SD=.021, p<<.01) and old facial expressions (mean=.382, SD=.020, p<<.01).

**A significant interaction was observed** (F (2, 266) = 4.347, p=.014) **between participants' age and gender.** Bonferroni's post hoc tests were performed for each single factor (participants' age and gender). These tests revealed that:

a) Concerning participants' gender: old male participants (mean=.388, SD=.037) showed better performances compared to old female participants (mean=.257, SD=.035).
b) Concerning participants' age: old female participants (mean=.257, SD=.035) showed worse performances compared to young females (mean=.464, SD=.033, p<<.01) and middle-aged female participants (mean=.483, SD=.034, p<<.01).
**A significant interaction was observed** (F (4, 532) = 2.962, p=.019) **between age of stimuli and participants' age.** Bonferroni's post hoc tests were performed for each single factor (participants' age and age of stimuli). These tests revealed that:
a) Concerning participants' age: young faces were worse decoded by old participants (mean=.376, SD=.039) compared to middle-aged (mean=.557, SD=.039, p=.004) and young participants (mean=.562, SD=.039, p=.003); middle-aged faces were better decoded by middle-aged participants (mean=.423, SD=.036) rather than by old participants (mean=.282, SD=.036, p=.018); old faces were better decoded by young participant (mean=.484, SD=.035) compared to middle-aged (mean=.354, SD=.035, p=.026) and old participants (mean=.309, SD=.035, p=.002).
b) Concerning age of stimuli: young participants worse decoded middle-aged facial expressions (mean=.355, SD=.035) compared to young (mean=.562, SD=.039, p<<.01) and old faces (mean=.484, SD=.035, p=.015); middle-aged participants better decoded young facial expressions (mean=.557, SD=.039) compared to middle-aged (mean=.423, SD=.036, p=.013) and old faces (mean=.354, SD=.035, p<<.01).
**A significant interaction emerged** (F (2, 532) = 14.765, p<<.01) **between stimuli's age and the type of stimulus (naturalistic vs synthetic) administered.** Bonferroni's post hoc tests were performed for each single factor (type and age of stimuli). These tests revealed that:
a) Concerning the type of stimulus: young faces were better recognized by the groups administered with synthetic facial expressions (mean=.559, SD=.032) compared to the groups required to recognize naturalistic faces (mean=.438, SD=.032, p=.007) and the same occurred for middle-aged faces which were better recognized by the groups administered with synthetic facial expressions (mean=.472, SD=.029) compared to the groups required to recognize naturalistic faces (mean=.235, SD=.029, p<<.01).
b) Concerning age of stimuli: the groups required to decode naturalistic facial expressions worse decoded middle-aged facial expressions (mean=.235, SD=.029) compared to young (mean=.438, SD=.032, p<<.01) and old ones (mean=.408, SD=.029, p<<.01). The groups required to decode synthetic facial expressions worse decoded old facial expressions (mean=.472, SD=.029) compared to young (mean=.559, SD=.032, p<<.01) and middle-aged ones (mean=.472, SD=.029, p=.006).
**A significant interaction emerged** (F (1, 266) = 42.012, p<<.01) **between stimuli's gender and the type of stimulus (naturalistic vs synthetic) administered.** Bonferroni's post hoc tests were performed for each single factor (type and gender of stimuli). These tests revealed that:
a) Concerning the type of stimulus: female facial expressions were better decoded by the groups required to decode synthetic facial expressions (mean=.513, SD=.024) compared to the groups required to label naturalistic facial expressions (mean=.278, SD=.025, p<<.01).
b) Concerning gender of stimuli: the groups required to decode naturalistic facial expressions better decoded male facial expressions (mean=.443, SD=.026) compared to female ones (mean=.278, SD=.025, p<<.01) while the groups required to decode synthetic facial expressions which better decoded female facial expressions (mean=.513, SD=.024) compared to male ones (mean=.413, SD=.025, p=.001)
**A significant interaction emerged** (F (2, 532) = 25.724, p<<.01) **between stimuli's age and gender**. Bonferroni's post hoc tests were performed for each single factor (age and gender of stimuli). These tests revealed that:
a) Concerning stimuli's age: when decoding female facial expressions, participants better recognized surprise conveyed by middle-aged (mean=.443, SD=.029) compared to old (mean=.328, SD=.028, p=.011). When decoding males, participants better decoded young faces (mean=.583, SD=.029) compared to middle-aged (mean=.264, SD=.025, p<<.01) and old ones (mean=.437, SD=.027, p<<.01).
b) Concerning stimuli's gender: when decoding young faces participants better recognized male (mean=.583, SD=.029) rather than females (mean=.414, SD=.028, p<<.01), when decoding middle-aged faces participants better decoded females (mean=.443, SD=.029) rather than male faces (mean=.264, SD=.025, p<<.01), when decoding old faces, participants better decoded males (mean=.437, SD=.027) rather than female faces (mean=.328, SD=.028, p=.004).

*Neutrality*

**Significant effects of participants' gender emerged** (F (1,266) = 8.770, p=.003). Bonferroni post hoc tests revealed that this was due to female participants (mean=.555, SD=.023) who showed better decoding performances compared to male participants (mean=.461, SD=.021, p=.003). **Significant effects of participants' age emerged** (F (2,266) = 8.912, p<<.01). Bonferroni post hoc tests revealed that this was due to young participants (mean=.601, SD=.027) who showed better decoding performances compared to middle-aged (mean=.452, SD=.027, p<<.01) and old participants (mean=.470, SD=.028, p=.003). **Significant effects of the type of stimulus emerged** (F (1,266) = 48.208, p<<.01). Bonferroni post hoc tests revealed that this was due to synthetic stimuli (mean=.617, SD=.022) were better decoded compared to naturalistic ones (mean=.398, SD=.023, p<<.01). **Significant effects of the gender of stimuli were observed** (F (1, 266) = 10.459, p=.001). Bonferroni post hoc tests revealed that this was due to male facial expressions (mean=.544, SD=.019) which were better recognized compared to female facial expressions (mean=.472, SD=.020, p=.001). **A significant interaction was observed** (F (2, 266) = 3.368, p=.036) **between participants' age and the type of stimulus administered**. Bonferroni's post hoc tests were performed for each single factor (participants' age and stimulus type). These tests revealed that:
a) Concerning participants' age: among the groups who were administered naturalistic stimuli, young participants (mean=.460, SD=.039) showed better recognition performances compared to middle-aged participants (mean=.315, SD=.039 p=.027), the same occurred among the groups who were administered synthetic facial expressions, with young participants (mean=.742, SD=.038) showing better performances compared to middle-aged (mean=.589, SD=.038, p=.014) and old participants (mean=.522, SD=.039, p<<.01).

b) Concerning the stimulus type: young participants who were administered synthetic stimuli (mean=.742, SD=.038) showed better performances compared to the group of young participants who were administered naturalistic stimuli (mean=.460, SD=.039, p<<.01); the same happened for the group of middle-aged participants who were administered synthetic stimuli (mean=.589, SD=.038) which showed better performances compared to the group of middle-aged participants who were administered naturalistic stimuli (mean=.315, SD=.039, p<<.01).

**A significant interaction emerged** (F (2, 532) = 29.551, p=<<.01) **between stimuli's age and the type of stimulus (naturalistic vs synthetic) administered.** Bonferroni's post hoc tests were performed for each single factor (type and age of stimuli). These tests revealed that:

a) Concerning the type of stimulus: young faces were better recognized by the groups administered with synthetic facial expressions (mean=.660, SD=.029) compared to the groups required to recognize naturalistic faces (mean=.381, SD=.030, p<<.01). Middle-aged faces were better recognized by the groups administered with synthetic facial expressions (mean=.691, SD=.029) compared to the groups required to recognize naturalistic faces (mean=.324, SD=.029, p<<.01).

b) Concerning age of stimuli: the groups required to decode naturalistic facial expressions better decoded old facial expressions (mean=.489, SD=.031) compared to young (mean=.381, SD=.030, p=.009) and middle-aged ones (mean=.324, SD=.029, p<<.01) while the groups required to decode synthetic facial expressions worse decoded old facial expressions (mean=.501, SD=.031) compared to young (mean=.660, SD=.029, p<<.01) and middle-aged ones (mean=.691, SD=.029, p<<.01).

**A significant interaction emerged** (F (2, 532) = 14.368, p<<.01) **between stimuli's age and gender**. Bonferroni's post hoc tests were performed for each single factor (age and gender of stimuli). These tests revealed that:

a) Concerning stimuli's age: when decoding female facial expressions, participants better recognized neutrality conveyed by middle-aged (mean=.529, SD=.028) compared to young (mean=.408, SD=.028, p=.002). When decoding males, participants better decoded young faces (mean=.633, SD=.027) compared to middle-aged (mean=.486, SD=.026, p<<.01) and old ones (mean=.512, SD=.030, p =.004).

b) Concerning stimuli's gender: when decoding young faces participants better recognized male (mean=.633, SD=.027) rather than female (mean=.408, SD=.028, p<<.01) facial expressions.